\documentclass[prd,preprint,amsmath,amssymb,nofootinbib,eqsecnum]{revtex4-1}
\usepackage{latexsym,graphics,color,epsfig,hyperref,mathrsfs}

\newcommand{\ie}{{i.e.}}

\newcommand{\viz}{{viz.}}
\newcommand{\wrt}{with respect to}
\newcommand{\lhs}{left-hand side}

\newcommand{\rhs}{right-hand side}

\newcommand{\be}{\begin{equation}}
\newcommand{\ee}{\end{equation}}
\newcommand{\bse}{\begin{subequations}}
\newcommand{\ese}{\end{subequations}}

\newcommand{\bcf}{\begin{center}\begin{figure}}
\newcommand{\ecf}{\end{figure}\end{center}}

\newcommand{\bct}{\begin{center}\begin{table}}
\newcommand{\ect}{\end{table}\end{center}}

\newcommand{\eq}[1]{(\ref{eq:#1})}

\newcommand{\eqs}[2]{(\ref{eq:#1}) and~(\ref{eq:#2})}

\newcommand{\sect}[1]{section~\ref{sec:#1}}

\newcommand{\app}[1]{appendix~\ref{app:#1}}

\newcommand{\D}{d}
\newcommand{\Int}[1]{\int \!\! d^{\D} \! #1 \,}

\newcommand{\DD}[1]{\delta^{(#1)}}

\newcommand{\measure}[1]{\bigl[ d #1 \bigr]}

\newcommand{\der}[2]{\frac{d #1}{d #2}}

\newcommand{\pder}[2]{\frac{\partial #1}{\partial #2}}

\newcommand{\fder}[2]{\frac{\delta #1}{\delta #2}}

\newcommand{\partialprod}[3]{D_{\underline{#1}_{#2}^{#3}}}

\newcommand{\Or}{\mathrm{O}}
\newcommand{\order}[1]{\Or \bigl( #1 \bigr)}
\newcommand{\dsorder}[1]{\Or (#1) }

\newcommand{\hf}{\frac{1}{2}}
\newcommand{\smallhf}{{\textstyle \frac{1}{2}}}

\newcommand{\one}{1\!\mathrm{l}}

\newcommand{\Tr}{\mathrm{Tr}\,}

\newcommand{\SCT}{\mathcal{K}}
\newcommand{\Dil}{\mathcal{D}}

\newcommand{\rot}[1]{L_{#1}}
\newcommand{\dil}[1]{D^{(#1)}}
\newcommand{\sct}[2]{{K^{(#1)}}_{#2}}

\newcommand{\quasi}{\mathcal{O}}
\newcommand{\subquasi}{\mathscr{O}}
\newcommand{\qsource}{J}
\newcommand{\subqsource}{\mathscr{J}}
\newcommand{\fsource}{J}
\newcommand{\emt}{T}
\newcommand{\emtop}{\hat{\emt}}
\newcommand{\emtsource}{\tau}

\newcommand{\fp}[1]{#1}

\newcommand{\acom}[2]{\bigl\{#1,#2\bigr\}}
\newcommand{\comm}[2]{\bigl[#1,#2\bigr]}

\newcommand{\pf}{\mathcal{Z}}
\newcommand{\field}{\phi}
\newcommand{\dfield}{\varphi}

\newcommand{\cutoff}{K}
\newcommand{\ep}{\mathcal{G}}

\newcommand{\Stot}{S}

\newcommand{\dual}{\mathcal{W}}
\newcommand{\R}{R}

\newcommand{\Ltot}{\hat{L}}

\newcommand{\classical}{\mathcal{C}}

\newcommand{\jop}{\hat{F}}

\newcommand{\hepth}[1]{hep-th/#1}

\begin{document}

\title{Wilsonian Ward Identities}
\author{Oliver J.~Rosten}
\email{oliver.rosten@gmail.com}
\affiliation{Unaffiliated}

\begin{abstract}
	For conformal field theories, it is shown how the Ward identity corresponding to dilatation invariance
	arises in a Wilsonian setting. In so doing, several points which are opaque in textbook treatments are
	clarified. Exploiting the fact that the Exact Renormalization Group furnishes a representation of the 	
	conformal algebra allows dilatation invariance to be stated directly as a property of the action, despite
	the presence of a regulator. This obviates the need for formal statements that conformal 
	invariance is recovered once the regulator is removed. Furthermore, the proper subset of conformal 
	primary fields for which the Ward identity holds is identified for all dimensionalities.
	
	\vspace{8ex}

	\begin{center}
		\large \em{In Varietate Concordia}
	\end{center}
\end{abstract}

\maketitle

\section{Introduction}

In classical field theory, Noether's theorem states that continuous symmetries of the action correspond to conserved currents. If such a symmetry is preserved at the quantum level then it implies relationships satisfied by the correlation functions: the Ward identities.

For the purposes of this paper, we wish to state symmetry constraints in a functional form. A simple example follows. Working in Euclidean space, suppose that our action depends on a scalar field, $\dfield$, \viz\ $\Stot[\dfield]$. If the action is translationally invariant then we can express this according to:
\be
	\partial_\mu \dfield \cdot \fder{}{\dfield} \Stot[\dfield] = 
	\Int{x} \partial_\mu \dfield(x) \fder{}{\dfield(x)} \Stot[\dfield]
	=
	0
.
\ee
Adapting standard derivations of Noether's theorem to our functional context, we insert a position-dependent parameter, $\epsilon(x)$, inside the relevant symmetry generator(s). For example:
\be
	\epsilon_\mu \partial_\mu \dfield \cdot \fder{}{\dfield}
	=
	\Int{x} \epsilon_\mu(x) 
	\partial_\mu \dfield(x) \fder{}{\dfield(x)}
.
\ee
This deformed generator will annihilate a translationally invariant action for constant $\epsilon_\mu$. For general epsilon we therefore expect
\be
	\epsilon_\mu \partial_\mu \dfield \cdot \fder{}{\dfield} \Stot[\dfield] 
	= \Int{x} \emt_{\alpha \mu} \partial_\alpha \epsilon_\mu(x)
	= -\Int{x} \epsilon_\mu(x) \partial_\alpha \emt_{\alpha \mu}
\ee
where, in this example, $T_{\alpha \mu}$ is the conserved current associated with translation invariance \ie\ the energy-momentum tensor.%
\footnote{At least up to terms which are exactly conserved.} 
That it is conserved follows from recalling the equations of motion: $\delta \Stot / \delta \dfield = 0$, whereupon it is apparent that, on the equations of motion,
\be
	\partial_\alpha \emt_{\alpha\mu} = 0
.
\ee

To make the transition to quantum field theory we will, for simplicity, consider coupling the fundamental field to the action via a source (treating composite operators in this manner is more subtle and will be discussed later):
\be
	\Stot[\dfield, \qsource] = \Stot[\dfield] - \qsource \cdot \dfield
.
\ee
Sticking with the example above, consider
\be
	\int\measure{\dfield}
	\biggl(
		\epsilon_\mu \partial_\mu \dfield \cdot \fder{}{\dfield}
		+
		\epsilon_\mu \partial_\mu \qsource \cdot \fder{}{\qsource}
	\biggr)
	e^{-\Stot[\dfield, \qsource]}
	=
	\int \measure{\dfield}
	e^{-\Stot[\dfield, \qsource]}
	\epsilon_\mu \cdot \partial_\alpha
	\bigl(
	    \emt_{\alpha\mu}
	  +\delta_{\alpha\mu} \qsource \times \dfield
	\bigr)
\label{eq:pre-WID}
\ee
where, in contrast to $\qsource \cdot \dfield$, $\qsource \times \dfield$ emphasises the absence of an integral.

Since the measure is translationally invariant, the first term on the \lhs\ is seen to be generated by the (infinitesimal) field redefinition
\be
	\dfield \rightarrow \dfield + \epsilon_\mu \partial_\mu \dfield
\ee
and so can be discarded. Note that, as should be the case, this field redefinition can be considered to arise from $x_\mu \rightarrow x_\mu + \epsilon_\mu$.
Focussing on the surviving terms and performing a functional derivative \wrt\ $\epsilon_\mu(x)$ yields:
\be
	\partial_\mu \qsource \times \fder{}{\qsource}
	\int \measure{\dfield}
	e^{-\Stot[\dfield, \qsource]}
	=
	\int \measure{\dfield}
	e^{-\Stot[\dfield, \qsource]}
	\partial_\alpha \emt_{\alpha\mu}
	+
	\partial_\mu 
	\biggl(
		\qsource \times \fder{}{\qsource}
		\int \measure{\dfield}
		e^{-\Stot[\dfield, \qsource]}
	\biggr)
,
\ee
 which reduces to the Ward identity
 \be
 	-\qsource \times \partial_\mu \fder{}{\qsource}
	\int \measure{\dfield}
	e^{-\Stot[\dfield, \qsource]}
	=
	\int \measure{\dfield}
	e^{-\Stot[\dfield, \qsource]}
	\partial_\alpha \emt_{\alpha\mu}
.
 \ee
 
 So far, all of this is standard. However, henceforth we wish to consider the case where conformal symmetry is realized in a non-linear manner. The motivation is as follows. In the above example, when making the transition from classical to quantum field theory, we blithely ignored that it may be necessary to regularize the path integral. We can get away with this sleight of hand since the path integral measure is translationally invariant and there is no difficulty regularizing the action in a way which respects
translation invariance. In contrast, for dilatation invariance matters are rather more subtle.

Given a scale-invariant classical action, in general we expect a regularization scheme to break this invariance. Specifically, classical dilatation invariance may be expressed as:
\be
	\dil{\delta} \dfield \cdot \fder{}{\dfield} \Stot[\dfield] 
	= 0
,
\ee
where $\delta$ is the classical scaling dimension and
\be
	\dil{\delta} = x_\mu \partial_\mu + \delta
.
\ee
As an example, it is straightforward to check that, in $\D=4$,
\be
	\dil{\delta_0} \dfield \cdot \fder{}{\dfield}
	\Int{x}
	\bigl(
		\smallhf \partial_\mu \dfield \, \partial_\mu \dfield + \dfield^4
	\bigr)
	=
	0
,
\ee
with
\be
	\delta_0 = \frac{\D-2}{2}
.
\label{eq:delta_0}
\ee

Now consider introducing cutoff regularization, via an ultra-violet cutoff function, $\cutoff(x,y)$. Inserting this into the kinetic term of the action above spoils dilatation invariance, at least at the classical level. However, in the full quantum field theory dilatation invariance is a property of the correlation functions and not the action, \emph{per se}. Therefore, it is conceivable that a regularized action, though not classically dilatation invariant, may ultimately yield a scale invariant quantum theory. (Of course, as is well known, for the quartic example given above the quantum theory is after all not scale invariant.)

This discussion may give the impression that the only diagnostic for quantum scale (or conformal) invariance comes from the correlation functions. However, the Exact Renormalization Group (ERG) provides a framework in which dilatation invariance is realized as a non-linear constraint on the Wilsonian effective action. The key point is that this non-linearity enables us to formulate a condition for dilatation invariance as a property of a regularized action and hence directly in the presence of a cutoff function. This point is central to~\cite{Representations} and has been recently emphasised in~\cite{Sonoda:ERG/CFT}.

The purpose of this paper is to understand how the Ward identities arise within the ERG representation of the conformal algebra. As a bonus, a further point that is rather opaque in a more standard treatment will be clarified. Notably, in~\cite{YellowPages}, it is carefully demonstrated how the dilatation Ward identity follows for correlation functions of the fundamental field. However, in general $\D$, no concrete statement is made about the other fields for which the Ward identity also holds; moreover, in $\D=2$ it is simply asserted that it has already been shown that the Ward identity holds for all Virasoro%
\footnote{In this paper, we shall refer to fields for which the correlation functions exhibit global conformal covariance as  conformal primary fields; in $\D=2$ these are typically called quasi-primary, with the Virasoro primary fields being those whose correlation functions posses local conformal covariance.}
primary fields!
(Subsequently, however, an independent derivation of the Ward Identity is supplied for $\D=2$.)
In this paper, a precise statement will be demonstrated for scalar conformal primary fields in scalar field theory: the Ward identity, in any $\D$, holds for the subset of such fields which, in the limit that the regularization is removed (\ie\ the classical limit), have no derivative terms. An analogous statement holds for fermionic fields.

Since, in this paper, the ERG is our tool of choice it will be introduced in \sect{ERG}. However, rather than providing a Wilsonian motivation for this framework~\cite{Wilson}, we instead follow~\cite{Representations} and introduce it as furnishing a particular representation of the conformal algebra. This viewpoint is more sympathetic to the thrust of the paper. After assembling some basic properties of the ERG, the sought-for Ward identity follows rather directly in \sect{Derivation}, by constructing appropriate solutions to the ERG equation. Following an example in~\sect{Example}, our conclusions are presented.

\section{The Exact RG}
\label{sec:ERG}

\subsection{Motivation and Equations}

Building on the examples presented in the introduction, in this section we will make more precise the notion of the ERG realizing dilatation invariance as a non-linear constraint on the Wilsonian effective action.
 
 Anticipating that the ERG equation has a partner which enforces invariance under special conformal transformations, let us start by constructing a `classical' functional representation of the conformal algebra:
 \be
 	\partial_\mu \dfield \cdot \fder{}{\dfield},
	\qquad
	\rot{\mu\nu} \dfield \cdot \fder{}{\dfield},
	\qquad
	\dil{\delta} \dfield \cdot \fder{}{\dfield},
	\qquad
	\sct{\delta}{\mu} \dfield \cdot \fder{}{\dfield}
,
\label{eq:ClassicalAlgebra}
 \ee
where
\be
	\rot{\mu\nu} = x_\mu \partial_\nu - x_\nu \partial_\mu,
	\qquad
	\dil{\delta} = x_\mu \partial_\mu + \delta,
	\qquad
	\sct{\delta}{\mu} = 2x_\mu \bigl(x_\alpha \partial_\alpha + \delta \bigr)	 -x^2 \partial_\mu
.
\label{eq:CFA-Classical}
\ee
It can be confirmed that the generators of~\eq{ClassicalAlgebra} satisfy the (Euclidean) conformal algebra, as required.

Our immediate aim is to understand how to state conformal invariance as a property of (interacting) conformal field theories. One option is to focus directly on the correlation functions. Again taking $\qsource$ to couple to the fundamental field, we can express conformal invariance as a property of the Schwinger functional, $\dual[\qsource]$. In particular, we can furnish a representation of the conformal algebra in which we swap out $\dfield$ for $\qsource$ in~\eq{CFA-Classical} whereupon, bearing in mind that $\qsource$ has scaling dimension $\D-\delta$, we have:
\be
	\partial_\mu \qsource \cdot \fder{\dual[\qsource]}{\qsource} = 0,
	\qquad
	\rot{\mu\nu} \qsource \cdot \fder{\dual[\qsource]}{\qsource} = 0,
	\qquad
	\dil{\D-\delta} \qsource \cdot \fder{\dual[\qsource]}{\qsource} = 0,
	\qquad
	\sct{\D-\delta}{\mu} \qsource \cdot \fder{\dual[\qsource]}{\qsource} = 0
.
\ee
There are two points to note. First, $\dual[\qsource]$ is an inherently non-local object.
Secondly, the scaling dimension of the fundamental field, $\delta$, is undetermined.

One approach to determine $\delta$ is to return to a local formulation of the theory. However, if we do so,
then we must accept a more complicated representation of the conformal algebra than furnished by~\eq{CFA-Classical}. Indeed, this is precisely what is achieved by fixed-point version of the Exact Renormalization Group (ERG). In this approach, the functional generators for translations and rotations are untouched. However, the dilatation and special conformal generators are modified as follows. 

Recalling the cutoff function, $\cutoff$, mentioned in the introduction, let us introduce some derived quantities:
\begin{subequations}
\begin{align}
	\bigl(
		\D + x\cdot \partial_x + y \cdot \partial_y
	\bigr)
	\cutoff (x,y)
	& = \partial^2_x (x,y) G(x,y)
,
\label{eq:G}
\\
	G_\mu(x,y)
	& = (x + y)_\mu G (x,y)
,
\\
	\ep = \ep_0 \cdot \cutoff
\label{eq:ep}
\end{align}
\end{subequations}
where
\be
	-\partial^2 \ep_0 = \one
.
\label{eq:Green}
\ee
Note that~\eqs{G}{ep} may be more intuitive in momentum space where, overloading notation, they read
$ G(p^2) = 2 \,d \cutoff(p^2) / d p^2$ and $\ep(p^2) = \cutoff(p^2)/p^2$. For future reference note that, in a derivative expansion (equivalently a Taylor expansion $p^2$), the lowest order contribution to $G(p^2)$ is
the dimensionless number $2 \cutoff'(0)$. An object exhibiting an all-orders derivative expansion is referred to as quasi-local.

Utilizing~\eq{delta_0} and defining $\eta$ according to
\be
	\fp{\eta} = \fp{\delta} - \delta_0
,
\ee
the generators of dilatations and special conformal transformations respectively take the form:
\begin{subequations}
\begin{align}
	\Dil 
	& = 
	\dil{\delta} \dfield \cdot \fder{}{\dfield}
	+ \dfield \cdot \ep^{-1} \cdot G \cdot \fder{}{\dfield}
	+ \hf \fder{}{\dfield} \cdot G \cdot \fder{}{\dfield}
,
\label{eq:Dil-ERG}
\\
	\SCT_\mu
	&=
	\sct{\delta}{\mu} \dfield \cdot \fder{}{\dfield}
	+ \dfield \cdot \ep^{-1} \cdot G_\mu \cdot \fder{}{\dfield}
	+ \hf \fder{}{\dfield} \cdot G_\mu \cdot \fder{}{\dfield}
	- \fp{\eta} \, \partial_\alpha \dfield \cdot \cutoff^{-1} \cdot G \cdot \fder{}{\dfield}
\label{eq:SCT-ERG}
.
\end{align}
\end{subequations}
As shown in~\cite{Representations}, these generators satisfy the conformal algebra. Interestingly, the generators annihilate not the action, but rather $e^{-\Stot}$. These conditions can be re-expressed as non-linear equations for the action, which self-consistently determine $\delta$ (at least in principle). The condition corresponding to dilatation invariance is what is known as the ERG equation, in the canonical form of~\cite{Ball}.

It will prove profitable to spend a few moments understanding the limit in which these generators reduce to their classical forms. This is most readily seen by transforming to dimensionful variables and working in momentum space; this procedure is discussed at length in~\cite{Fundamentals} and is revisited in \app{Details}. Denoting the energy scale  by $\Lambda$, the transformed field, $\field(p)$, is accompanied by a factor of $\Lambda^{\D-\fp{\delta}}$ from which it follows that  $\delta / \delta \field(p)$ picks up a factor of $\Lambda^{\delta}$.  Bearing in mind that a momentum integral acquires a factor of $\Lambda^{-\D}$ after transferring to dimensionful units, it is clear that the classical contributions to~\eqs{Dil-ERG}{SCT-ERG} survive the classical limit $\Lambda \rightarrow \infty$. What of the remaining terms?

Consider the second contribution to $\Dil$. Compared to a raw $\field \cdot \delta / \delta \field$---which we have just seen survives the limit--- this term has an extra contribution which, at leading order in momentum, goes like $ \cutoff(0) \cutoff'(0) p^2/\Lambda^2 $. Manifestly, this is sub-leading in the $\Lambda \rightarrow \infty$ limit. As for the final contribution to $\Dil$, this goes like $\Lambda^{2\delta - d} = \Lambda^{\eta - 2}$. This is sub-leading so long as $\eta < 2$, which we now assume to be the case. The special conformal generator, \eq{SCT-ERG}, can be similarly treated.

\subsection{Properties of Solutions}

There are several properties of solutions to the ERG equation that will be required.  For what follows, we shall consider a solution, $\fp{\Stot}$, that also solves the ERG equation's special conformal partner and, as such, corresponds to a conformal field theory. For such theories, there always exists a particular pair of scalar conformal primary fields, which we shall specify momentarily. Recall that a scalar conformal primary field of scaling dimension $\Delta$ is defined according to:
\be
	\Dil \quasi^{(\Delta)} = \dil{\Delta} \quasi^{(\Delta)}
	\qquad
	\SCT_\mu \quasi^{(\Delta)} = \sct{\Delta}{\mu} \quasi^{(\Delta)}
.
\ee

Introducing a function, $\R$, which is most readily expressed in momentum space
\be
	\R(p^2) = p^{2(\fp{\eta}/2-1)} \cutoff(p^2)
	\int_0^{p^2} d q^2
	q^{-2(\fp{\eta}/2)}
	\der{}{q^2} 
	\frac{1}{\cutoff(q^2)}
	\qquad
	\fp{\eta} < 2
,
\label{eq:rho}
\ee
the promised pair of conformal primary fields are given by~\cite{Representations,HO-Remarks}:
\begin{subequations}
\begin{align}
	\quasi^{(\delta)} e^{-\fp{\Stot}} & = 
	\biggl(
		\cutoff^{-1} \cdot \dfield + \R \cdot \fder{}{\dfield}
	\biggr)
	e^{-\fp{\Stot}}
,
\\
	\quasi^{(\D - \delta)} e^{-\fp{\Stot}} & = 
	-
	\fder{}{\dfield} \cdot \cutoff \, e^{-\fp{\Stot}} 
.
\label{eq:O^(d-delta)}
\end{align}
\end{subequations}
To provide some intuition, note that $\quasi^{(\delta)}$ is morally the fundamental field: it is easy to show that, in the limit that the regularization is removed, it simply reduces to $\dfield$~\cite{Fundamentals}. On the other hand, $\quasi^{(\D -\delta)}$ is the redundant field which, in the classical limit, defines the equations of motion.

From~\eq{O^(d-delta)}, we can derive a `conformal operator', which we define such that
\be
	\comm{\Dil}{\hat{\quasi}^{(\Delta)}} = \dil{\Delta} \hat{\quasi}^{(\Delta)}
	\qquad
	\comm{\SCT_\mu}{\hat{\quasi}^{(\Delta)}} = \sct{\Delta}{\mu} \hat{\quasi}^{(\Delta)}
.
\ee
Specifically,
\be
	\hat{\quasi}^{(\D - \delta)}
	=
	\fder{\fp{\Stot}}{\dfield} \cdot \cutoff 
	-\fder{}{\dfield} \cdot \cutoff 
.
\ee
Using these ingredients, we may now build the trace of the energy-momentum tensor:
\be
	\emt = -\fp{\delta} \, \hat{\quasi}^{(\D - \delta)} \times {\quasi}^{(\delta)}
.
\ee
It will also prove fruitful to construct an operator corresponding to $\emt$:
\be
	\hat{\emt} = 
	\delta \,
	\fder{}{\dfield} \cdot \cutoff \times 
	\biggl(
		\cutoff^{-1} \cdot \dfield + \R \cdot \fder{}{\dfield}
	\biggr)
,
\ee
which satisfies
\be
	\hat{\emt} e^{-\fp{\Stot}} = \emt  e^{-\fp{\Stot}}
.
\label{eq:emtop}
\ee
Again, some intuition is provided by the classical limit where we see that, up to a possible vacuum term, the trace of the energy-momentum tensor is given by $T \sim -\fp{\delta} \, \dfield \times \delta \fp{\Stot} / \delta \dfield$. Actually, the vacuum contribution to $\hat{\emt}$ is divergent; this can be regularized by point-splitting and could, at least in principle, cancel out in $\emt$. This issue will be glossed over for now, since vacuum terms are not relevant to this paper; however, it will be dealt with properly in a forthcoming paper.

\subsection{Coupling Sources}

Computing correlation functions in the Wilsonian setting is naturally done by coupling sources to the action and supplementing the ERG equation (and its partner) with appropriate terms. Given a source, $\qsource$, which we understand to couple to some $\quasi^{(\Delta)}$, the dilatation generator, \eqref{eq:Dil-ERG} becomes:
\be
	\Dil_{\qsource} 
	= 
	  \dil{\D-\Delta} \qsource \cdot \fder{}{\qsource}
	+ \dil{\delta} \dfield \cdot \fder{}{\dfield}
	+ \dfield \cdot \ep^{-1} \cdot G \cdot \fder{}{\dfield}
	+ \hf \fder{}{\dfield} \cdot G \cdot \fder{}{\dfield}
,
\label{eq:DilJ-ERG}
\ee
with similar modifications to~\eq{SCT-ERG}. This can be trivially extended to include additional sources.

The goal now is to solve for $\fp{\Stot}[\dfield, \qsource]$ given an appropriate boundary condition. The latter is perhaps most intuitively stated in dimensionful variables. Recalling the discussion under~\eqs{Dil-ERG}{SCT-ERG}, the boundary condition is given in the limit that $\Lambda \rightarrow \infty$, the classical limit:
\be
	\lim_{\Lambda \rightarrow \infty}
	\Stot_\Lambda[\field, \qsource]
	=
	\Stot_{\infty}[\field] - \qsource \cdot \quasi_{\infty}
.
\ee

For completeness, it is worth describing what happens if we stick with dimensionless variables, as this is the language in which conformal field theories are naturally phrased. In this context, we demand that 
$\fp{\Stot}[\dfield, \qsource]$ is classically dilatation invariant, up to terms which can be removed by a quasi-local field redefinition. Of course, this turns out to be equivalent to the above prescription and is tied up with the fact that the ERG equation itself corresponds to a particular field redefinition. We will explore this more thoroughly in \app{Details}.

Returning to dimensionless variables, the solution to the ERG equation that we are looking for is simply~\cite{Fundamentals}
\be
	\fp{\Stot}[\dfield, \qsource] = \fp{\Stot}[\dfield] - \qsource \cdot \quasi + \order{\qsource^2}
,
\ee
where $\quasi$ (from which we have dropped the explicit scaling dimension) satisfies the eigenvalue equation:
\be
	\biggl(
		\dil{\D-\Delta} \qsource \cdot \fder{}{\qsource}
		-\dil{\delta} \dfield \cdot \fder{}{\dfield}
		- \dfield \cdot \ep^{-1} \cdot G \cdot \fder{}{\dfield}
		+ \hf \fder{\fp{\Stot[\dfield]}}{\dfield} \cdot G \cdot \fder{}{\dfield}
		-\hf \fder{}{\dfield} \cdot G \cdot \fder{}{\dfield}
	\biggr)
	\qsource \cdot\quasi
	=
	0
.
\ee
Thus, if we knew $\fp{\Stot[\dfield]}$ then, in principle, we could determine $\quasi$ and $\Delta$.

However, beyond leading order in $\qsource$, matters are much more delicate since, in general, the issue of the renormalization of composite operators must be addressed. Exceptions are the fundamental field and also the trace of the energy-momentum tensor for both of which we can explicitly construct an exact, source-dependent solution of the fixed-point ERG equation. 
Specifically, if we take $\fsource$ to couple to $\quasi^{(\fp{\delta})}$, then there exists the following solution to the (source-dependent) ERG equation~\cite{Fundamentals}:
\be
	\fp{\Stot}[\dfield, \fsource]
	=
	\fp{\Stot}[\dfield]
	+
	\bigl(
		e^{\fsource \cdot \fp{\R} \cdot \delta/\delta \dfield}
		-1
	\bigr)
	\bigl(
		\fp{\Stot}[\dfield] - \smallhf \dfield \cdot \cutoff^{-1} \cdot \fp{\R}^{-1} \cdot \dfield
	\bigr)
.
\label{eq:S_J}
\ee
Similarly, taking $\emtsource$ to couple to the $\emt$, we have a source-dependent solution given by:
\be
	e^{-\fp{\Stot}[\dfield, \emtsource]}
	=
	e^{\emtsource \cdot \emtop}
	e^{-\fp{\Stot}[\dfield]}
.
\label{eq:S[phi,tau]}
\ee

Nevertheless, for the purposes of this paper, we will mostly take the sources to be infinitesimal. To consider multiple insertions of $\quasi$, we can couple $\quasi$ to not one but an arbitrary number of infinitesimal sources with disjoint support.

\section{Wilsonian Ward Identity}
\label{sec:WID}

\subsection{Derivation}
\label{sec:Derivation}

In the introduction, we presented a functional derivation of the Ward identity following from translational invariance. This was achieved using a classical representation of the translation generator. Turing to dilatations, rather than attempting to mimic this in the quantum setting, we instead show
how the associated Ward identity is encoded in solutions to the ERG equation. As such, we consider an additional source, $\emtsource$, which couples to the energy-momentum tensor. The dilatation generator~\eqref{eq:DilJ-ERG} becomes:
\be
	\Dil_{\qsource, \emtsource} = 
	\dil{0}\emtsource \cdot \fder{}{\emtsource}
	+ \Dil_{\qsource}
.
\ee

From the associated ERG equation, we
now seek a solution, $\fp{\Stot}[\dfield, \qsource, \emtsource]$, with the following property in the classical limit:
\be
	\fp{\Stot}[\dfield, \qsource, \emtsource]
	\sim
	\fp{\Stot}[\dfield] - \qsource \cdot \quasi - \emtsource \cdot \emt
.
\ee
Note the crucial point that, in the expansion of the \rhs, there is no $\order{\qsource \emtsource}$ contribution. Indeed, given that $\fp{\Stot}[\dfield, \qsource, \emtsource]$ is a solution to the ERG equation, we can, courtesy of the fact that $\emtsource$ has zero scaling dimension, construct a family of solutions
\be
	e^{-\fp{\Stot}_a[\dfield, \qsource, \emtsource]}
	=
	\biggl(
		1 + a \Int{x} \emtsource(x) \qsource(x) \fder{}{\qsource(x)}
	\biggr)
	e^{-\fp{\Stot}[\dfield, \qsource, \emtsource]}
\ee
for $-\infty < a < \infty$. This additional term will survive the classical limit,
producing a contribution
\[
	a \Int{x} \emtsource(x) \qsource(x) \quasi(x)
.
\]
However, for non-zero $a$ this spoils the boundary condition.

We now proceed with the construction of $\fp{\Stot}[\dfield, \qsource, \emtsource]$. To begin, consider just a solution to the sourceless ERG equation, $\fp{\Stot}[\dfield]$. To couple this to the trace of the energy-momentum tensor, recall that
\be
	\emt e^{-\fp{\Stot}[\dfield]}
	=
	\fp{\delta}
	\fder{}{\dfield} \cdot \cutoff
	\times
	\biggl(
		\cutoff^{-1} \cdot \dfield
		+\fp{\R} \cdot \fder{}{\dfield}
	\biggr)
	e^{-\fp{\Stot}[\dfield]}
,
\ee
whereupon we construct
\be
	\fp{\Stot}[\dfield, \emtsource] = \fp{\Stot}[\dfield] - \emtsource \cdot \emt + \order{\emtsource^2}
,
\ee
which solves the ERG equation with the appropriate boundary condition.

Next, suppose we have a source-dependent solution, $\fp{\Stot}[\dfield, \qsource]$. From this, we can construct a source-dependent version of the energy-momentum tensor, the trace of which is given by
\be
	\emt_\qsource e^{-\fp{\Stot}[\dfield, \qsource]}
	=
	\emtop e^{-\fp{\Stot}[\dfield, \qsource]}
	=
	\fp{\delta}
	\fder{}{\dfield} \cdot \cutoff
	\times
	\biggl(
		\cutoff^{-1} \cdot \dfield
		+\fp{\R} \cdot \fder{}{\dfield}
	\biggr)
	e^{-\fp{\Stot}[\dfield, \qsource]}
.
\label{eq:emt_j}
\ee
In turn, we can construct a Wilsonian effective action depending on both $\qsource$ and $\emtsource$:
\be
	e^{-\fp{\Stot}'[\dfield, \qsource, \emtsource]}
	=
	\bigl(
		1 + \emtsource \cdot \emtop 
	\bigr)
	e^{-\fp{\Stot}[\dfield, \qsource]}
	+ \order{\emtsource^2}
.
\label{eq:S'}
\ee
However, although $\fp{\Stot}'$ solves the ERG equation, it does not satisfy the desired boundary condition---which we have emphasised with the prime. Indeed, under the functional integral,
\be
	\int \measure{\dfield} e^{-\fp{\Stot}'[\dfield, \qsource, \emtsource]}
	= 
	\int \measure{\dfield} e^{-\fp{\Stot}[\dfield, \qsource]}
	+
	\order{\emtsource^2},
\ee
due to the total functional derivative in~\eq{emt_j} and so it is clear that $\fp{\Stot}'[\dfield, \qsource, \emtsource]$ does not generate correlation functions of $\quasi$ involving a single insertion of $\emt$.

Nevertheless, $\fp{\Stot}'$ is close to the object we seek:
\be
	e^{-\fp{\Stot}[\dfield, \qsource, \emtsource]}
	=
	\biggl(
		1 + \emtsource \cdot \emtop 
		- \Delta' \Int{x} \emtsource(x) \qsource(x) \fder{}{\qsource(x)}
	\biggr)
	e^{-\fp{\Stot}[\dfield, \qsource]}
	+ \order{\emtsource^2}
.
\label{eq:WID-prelim}
\ee
It now remains to show that, for a particular subset of the conformal fields, the boundary condition mentioned above fixes  $\Delta' = \Delta$.

Before demonstrating this, let us suppose that it is true for some subset of conformal primary fields,  $\{\subquasi\}$, and explore the consequences. A representative of this set will be taken to couple to a source $\subqsource$. Returning to~\eq{WID-prelim} and working to $\order{\tau}$ gives:
\be
	\fder{}{\tau}
	e^{-\fp{\Stot}[\dfield, \subqsource, \emtsource]}
	\biggr\vert_{\tau=0}
	=
	\emtop
	e^{-\fp{\Stot}[\dfield, \subqsource]}
	-
	\Delta \subqsource \times \fder{}{\subqsource}
	e^{-\fp{\Stot}[\dfield, \subqsource]}
.
\ee
Performing a functional integral, the first term in the \rhs\ vanishes by virtue of being a total functional derivative, as we see from~\eq{emt_j}. The remaining two terms are simply a statement of the dilatation Ward identity. Recall that this derivation is valid mostly for infinitesimal source; however, it is trivial to generalize to an arbitrary number of such sources, each with disjoint support:
\be
	\fder{}{\tau}
	\int \measure{\dfield} e^{-\fp{\Stot}[\dfield, \{\subqsource\}, \emtsource]}
	\biggr\vert_{\tau=0}
	=
	-\sum_i \Delta_i \subqsource_i \times \fder{}{\subqsource_i}
	\int \measure{\dfield}
	e^{-\fp{\Stot}[\dfield, \{\subqsource\}]}
,
\ee
recovering the general statement of the Ward identity, albeit in functional form.

Let us now return to showing that the boundary condition on the action fixes $\Delta' = \Delta$.
The essence of the derivation is simple. The terms at $\order{\emtsource}$ may be written:
\be
	\fder{}{\emtsource}
	e^{-\fp{\Stot}[\dfield, \qsource, \emtsource]}
	\biggr\vert_{\emtsource = 0 }
	=
	\biggl[
		\fp{\delta}
		\fder{}{\dfield} \cdot \cutoff
		\times
		\biggl(
			\cutoff^{-1} \cdot \dfield
			+\fp{\R} \cdot \fder{}{\dfield}
		\biggr)
		-
		\Delta'
		\qsource \times \fder{}{\qsource}
	\biggr]
	e^{-\fp{\Stot}[\dfield, \qsource]}
.
\ee
Narrowing to $\order{\qsource \emtsource}$ terms, we take the classical limit:
\be
	\qsource \cdot \fder{}{\qsource}
	\fder{}{\emtsource}
	e^{-\fp{\Stot}[\dfield, \qsource, \emtsource]}
	\biggr\vert_{\emtsource = 0}
	\sim
	\fp{\delta}
	\dfield \times \fder{\qsource \cdot \quasi}{\dfield}
	-\Delta'
	\qsource \times \quasi
	+ \order{\qsource^2}
.
\ee
We shall shortly show that there is a proper subset of the conformal primary fields denoted, as above by $\{\subquasi\}$, for which
\be
	\fp{\delta}
	\, \dfield \times \fder{\subqsource \cdot \subquasi}{\dfield}
	\sim
	\Delta \subqsource \times \subquasi
.
\label{eq:classical_limit}
\ee
Given this, the boundary condition we seek to impose is satisfied if $\Delta' = \Delta$. Note that the subset is indeed a proper one, since irrespective of whichever other conformal fields may exist, the three fields $\quasi^{(\fp{\delta})}$, $\quasi^{(\D-\fp{\delta})}$ and $\emt$ belong to $\{\quasi\}$, but only the first of these belongs to $\{\subquasi\}$.

Before proceeding to demonstrate~\eq{classical_limit} we pause for an example. Consider the Gaussian fixed-point, for which $\delta = (\D-2)/2$. Restricting ourselves to the conformal fields which are just powers of $\dfield$, we can label $\subquasi$ by an integer which counts the number of fields:
\be
	\subquasi^{(n)} \sim \dfield^n,
	\qquad
	\Delta^{(n)} = n \frac{\D-2}{2}
,
\ee
confirming that~\eq{classical_limit} holds in this case. Note that, in this example, $\quasi^{(\D-\fp{\delta})} \sim -\partial^2 \dfield$ with $\emt \sim \delta_0 \dfield \times \partial^2 \dfield$.

To demonstrate~\eq{classical_limit}, we will utilize the `conformal fixed-point equation' of~\cite{WEMT}.
This is a single, unintegrated equation that encodes both the ERG equation and its special conformal partner. To state the equation, let us take $\Ltot$ to belong to the equivalence class of objects that integrate to the action. Now define
\be
	\partialprod{\alpha}{j}{i}
	\equiv
	\Biggl\{
	\begin{array}{cl}
		\prod_{k=j}^i \partial_{\alpha_k} & i \geq j,
	\\
		1 & i < j
,
	\end{array}
\ee
from which we construct
\begin{multline}
	\classical_{\Ltot}(x)
	=
	\delta \dfield \times \fder{\Stot}{\dfield}
	-\D\Ltot
	+
	\sum_{i=1}^\infty
	\bigl[ \partialprod{\sigma}{1}{i}, x\cdot\partial \bigr] \dfield
	\times
	\pder{\Ltot}{(\partialprod{\sigma}{1}{i}\dfield)}
\\
	-\partial_\lambda
	\biggl(
		\hf
		\bigl(
			\delta_{\omega \lambda} \delta_{\rho\sigma}
			- 2\delta_{\omega \rho} \delta_{\sigma\lambda}
		\bigr)
		\sum_{i=2}^\infty
		\bigl[
			\bigl[
				\partialprod{\sigma}{1}{i}, x_\sigma
			\bigr],
			x_\rho \partial_\omega
		\bigr]
		\dfield
		\times
		\pder{\Ltot}{(\partialprod{\sigma}{1}{i} \dfield)}
	\biggr)
.
\end{multline}
As demonstrated in~\cite{WEMT}, this has the following properties:
\be
	\Int{x} \classical_{\Ltot}(x)
	=
	\dil{\delta} \dfield \cdot \fder{\Stot}{\dfield}
	,
	\qquad
	\Int{x} 2 x_\mu \classical_{\Ltot}(x)
	=
	\sct{\delta}{\mu} \dfield \cdot \fder{\Stot}{\dfield}
.
\ee
The symbol, $ \classical$ is indicative that the associated contributions have a classical origin and may potentially survive the classical limit.
With this in mind, and using the notation
\be
	\acom{G}{\one}(y,z;x) = G(y,x) \DD{\D}(x-z) + \DD{\D}(y-x) G(x,z)
,
\label{eq:acom}
\ee
the conformal fixed-point equation reads:
\begin{multline}
	\order{\partial^2}
	=
	\classical_{\Ltot} 
	+
	\hf \dfield \cdot \ep^{-1} \cdot \acom{G}{\one} \cdot \fder{\Stot}{\dfield}
	-\hf \fder{\Stot}{\dfield} \cdot G \times \fder{\Stot}{\dfield}
	+\hf \fder{}{\dfield} \cdot G \times \fder{\Stot}{\dfield}
\\
	+
	\partial_\lambda
	\biggl(
		\frac{\eta}{4} \partial_\lambda \dfield \cdot \cutoff^{-1} 
		\cdot \acom{G}{\one} \cdot \fder{\Stot}{\dfield}
	\biggr)
.
\label{eq:CERG}
\end{multline}
Note that the extra terms on the \rhs\ are sub-leading in the classical limit, as follows straightforwardly from the analysis under~\eqs{Dil-ERG}{SCT-ERG}.

In the presence of a source, we simply shift
\be
	\Ltot \rightarrow \Ltot_{\qsource} = \Ltot - \qsource \times \quasi
,
\ee
together with
\begin{multline}
	\classical_{\Ltot}(x)
	\rightarrow
	\classical_{\Ltot_{\qsource}}(x)
	=
	\classical_{\Ltot}(x)
	-
	\fp{\delta} \dfield \times \fder{\qsource \cdot \quasi}{\dfield}
	+\Delta \qsource \times \quasi
	-
	\sum_{i=1}^\infty
	\bigl[ \partialprod{\sigma}{1}{i}, x\cdot\partial \bigr] \dfield
	\times
	\pder{\qsource \times \quasi}{(\partialprod{\sigma}{1}{i}\dfield)}
\\
	+
	\partial_\lambda
	\biggl(
		\hf
		\bigl(
			\delta_{\omega \lambda} \delta_{\rho\sigma}
			- 2\delta_{\omega \rho} \delta_{\sigma\lambda}
		\bigr)
		\sum_{i=2}^\infty
		\bigl[
			\bigl[
				\partialprod{\sigma}{1}{i}, x_\sigma
			\bigr],
			x_\rho \partial_\omega
		\bigr]
		\dfield
		\times
		\pder{\qsource \times \quasi}{(\partialprod{\sigma}{1}{i} \dfield)}
	\biggr)	
.
\end{multline}

Focussing on the $\order{\qsource}$ terms we now take the classical limit. If we restrict $\{\quasi\}$ to $\{\subquasi\}$, where the latter are defined such that, in the classical limit
\be
	\pder{\subquasi}{(\partialprod{\sigma}{1}{i}\dfield)} \sim 0,
	\qquad i > 0
\ee
then it is apparent that
\be
	\order{\partial^2} \sim - \fp{\delta} \, \dfield \times \fder{\subqsource \cdot \subquasi}{\dfield}
	+
	\Delta \subqsource \times \subquasi
.
\ee
However, since all $\order{\partial^2}$ contributing to the \rhs\ are sub-leading, in the classical limit the \lhs\ must vanish leaving
\be
	\fp{\delta} \, \dfield \times \fder{\subqsource \cdot \subquasi}{\dfield}
	\sim
	\Delta \subqsource \times \subquasi
,
\ee
as was to be demonstrated. 

\subsection{Example}
\label{sec:Example}

In the special case of the fundamental field, we have at our disposal a solution to the ERG equation valid to all orders in the $\fsource$, equation~\eq{S_J}. This allows us to relax the restriction that the source is infinitesimal.
To begin, let us follow the recipe of the previous section. First,
we construct the solution $\fp{\Stot}'[\dfield, \fsource, \emtsource]$ which, recalling~\eq{S'}, satisfies
\be
	\fder{}{\emtsource} 
	e^{-\fp{\Stot}'[\dfield, \fsource, \emtsource]}
	\biggl\vert_{\emtsource = 0}
	=
	\emtop 
	e^{-\fp{\Stot}[\dfield, \fsource]}
.
\ee
Secondly, we construct a new solution which respects the desired boundary condition:
\be
	\fder{}{\emtsource} 
	e^{-\fp{\Stot}[\dfield, \fsource, \emtsource]}
	\biggl\vert_{\emtsource = 0}
	=
	\biggl(
		\emtop 
		-
		\fp{\delta}\fsource \times \fder{}{\fsource}
	\biggr)
	e^{-\fp{\Stot}[\dfield, \fsource]}
,
\ee
thereby arriving at a statement of the Ward identity.

However, we now exploit the expression for $\fp{\Stot}[\dfield, \fsource]$ to derive the Ward identity in a different way. The form of this solution, \eq{S_J}, suggests that we can readily construct a related solution which also couples to $\emt$:
\be
	\fp{\Stot}[\dfield, \fsource, \emtsource]
	=
	\fp{\Stot}[\dfield,\emtsource]
	+
	\bigl(
		e^{\fsource \cdot \fp{\R} \cdot \delta/\delta \dfield}
		-1
	\bigr)
	\bigl(
		\fp{\Stot}[\dfield,\emtsource] - \smallhf \dfield \cdot 				\cutoff^{-1} \cdot \fp{\R}^{-1} \cdot \dfield
	\bigr)
	+\order{\emtsource^2}
,
\label{eq:S_Jep}
\ee
with $\fp{\Stot}[\dfield,\emtsource]$ given by~\eq{S[phi,tau]}.
It can be easily checked that this is both a solution of the ERG equation and that  the $\order{\fsource \emtsource}$ term does not survive the classical limit. Therefore, it must be true that this solution encodes the Ward identity.

To demonstrate the latter, observe that we can write
\be
	e^{-\fp{\Stot}[\dfield, \fsource]}
	=
	\jop
	e^{-\fp{\Stot}[\dfield]}
,
\qquad
	e^{-\fp{\Stot}[\dfield, \fsource, \emtsource]}
	=
	\jop
	e^{-\fp{\Stot}[\dfield, \emtsource]}
\ee
where
\be
	\jop[\dfield, \fsource, \delta / \delta \dfield] \equiv 
	e^{-\hf \dfield \cdot \cutoff^{-1} \cdot \fp{\R}^{-1} \cdot \dfield}
	e^{\fsource \cdot \fp{\R} \cdot \delta /\delta \dfield}
	e^{\hf \dfield \cdot \cutoff^{-1} \cdot \fp{\R}^{-1} \cdot \dfield}
.
\ee
This operator obeys the following commutators:
\be
	\biggl[
		\cutoff^{-1} \cdot \dfield
		+\fp{\R} \cdot \fder{}{\dfield}
	\,
	,
	\,
		\jop
	\biggr]
	=
	0
,
\qquad
	\biggl[
		\fder{}{\dfield} \cdot \cutoff
	,
		\jop
	\biggr]
	=
	\fsource
	\jop
,
\ee
from which is follows that
\be
	\emtop e^{-\fp{\Stot}[\dfield, \fsource]}
	=
	\biggl\{
	\delta \qsource \jop 
	\biggl(
		\cutoff^{-1} \cdot \dfield
		+\fp{\R} \cdot \fder{}{\dfield}
	\biggr)
	+ \jop \, \emtop 
	\biggr\}
	e^{-\fp{\Stot}[\dfield, \fsource]}
.
\ee
Observing that
\be
	\jop
	\;
	\biggl(
		\cutoff^{-1} \cdot \dfield
		+\fp{\R} \cdot \fder{}{\dfield}
	\biggr)
	=
	\fder{}{\fsource}
	\jop
,
\ee
yields
\be
	\emtop e^{-\fp{\Stot}[\dfield, \fsource]}
	=
	\biggl(
		\delta \fsource \times \fder{}{\fsource} \jop
		+ \jop \, \emtop 
	\biggr)
	e^{-\fp{\Stot}[\dfield]}
	=
	\delta \fsource \times \fder{}{\fsource} e^{-\fp{\Stot}[\dfield, \fsource]}
	+
	\fder{}{\emtsource}
	e^{-\fp{\Stot}[\dfield, \fsource, \emtsource]}
	\biggr\vert_{\emtsource = 0}
\ee
confirming that~\eq{S_Jep} does indeed encode the Ward identity.

\section{Conclusion}

For conformal field theories, the Ward identity corresponding to dilatation invariance has been shown to be encoded in solutions of the ERG equation. A particular benefit of deriving the Ward identity in this way is that it avoids questions as to whether the introduction of a regulator potentially spoils conformal invariance: 
by deforming the classical representation of the conformal generators, conformal invariance can be stated directly as a property of a regularized Wilsonian effective action, without any reference to the (non-local) correlation functions. Furthermore, this derivation straightforwardly identifies the subset of conformal primary fields for which the Ward identity actually holds:  those fields which, in the limit that the regularization is removed, do not contain any derivatives.

It is also noteworthy that, contrary to more standard derivations, no explicit use is made of the dilatation current. Indeed, only the trace the energy-momentum tensor appears in our analysis, cementing the increasingly central role that this object plays in understanding the mathematical structure of the ERG.

\appendix

\section{The Classical Limit}
\label{app:Details}

To begin, consider the effects of performing a quasi-local infinitesimal field redefinition under the functional integral:
\be
	\dfield \rightarrow \dfield' = \dfield + \varepsilon \Psi
,
\ee
where $\Psi$ may depend on $\dfield$.
The action in terms of the new field satisfies:
\be
	\fp{\Stot}[\dfield, \qsource]
	=
	\fp{\Stot}[\dfield' - \varepsilon \Psi, \qsource]
	=
	\fp{\Stot}[\dfield', \qsource]
	-
	\varepsilon
	\Psi \cdot \fder{\fp{\Stot}[\dfield', \qsource]}{\dfield'}
	+
	\order{\varepsilon^2}
,
\ee
whereas the functional measure transforms according to
\be
	\measure{\dfield}
	=
	\measure{\dfield'}
	\biggl\vert
		\fder{\dfield}{\dfield'}
	\biggr\vert
.
\ee
Exploiting the formula $\det X = e^{\Tr \ln X}$ yields
\be
	\measure{\dfield}
	=
	\measure{\dfield'}
	\biggl(
		1 - \varepsilon \fder{}{\dfield'} \cdot \Psi
	\biggr)
	+
	\order{\varepsilon^2}
.
\ee
Putting everything together recovers the expected results: under an infinitesimal field redefinition, the action and measure together generate a total functional derivative. Specifically,
\be
	\measure{\dfield} e^{-\fp{\Stot}[\dfield, \qsource]}
	=
	\measure{\dfield'} 
	\biggl(
		1
		-
		\varepsilon
		\fder{}{\dfield'}
		\cdot
		\Psi
	\biggr)
	e^{-\fp{\Stot}[\dfield', \qsource]}
.
\ee

With this in mind, let us suppose that $\fp{\Stot}[\dfield, \qsource]$ solves the (source-dependent) ERG equation. Inside the action functional, consider performing a change of coordinates, corresponding to an infinitesimal dilatation:
\be
	x_\mu \rightarrow x'_\mu = (1+ \epsilon) x_\mu
.
\label{eq:coordinate_shift}
\ee
Recalling~\eq{DilJ-ERG}, this induces a change in the action which vanishes on account of the action satisfying the ERG equation:
\be
	e^{-\fp{\Stot}[\dfield, \qsource]}
	\rightarrow
	e^{-\fp{\Stot}'[\dfield, \qsource]}
	=
	\bigl(
		1 + \epsilon\, \Dil_{\qsource}
	\bigr)
	e^{-\fp{\Stot}[\dfield, \qsource]}
	=
	e^{-\fp{\Stot}[\dfield, \qsource]}
	+\order{\epsilon^2}
.
\label{eq:infinitesimalDil}
\ee
The key observation is that, up to a (divergent) vacuum term, $V$, we may write $\Dil$ as a total functional derivative, yielding
\be
	\pf'[\qsource] \equiv
	\int
	\measure{\dfield} 
	e^{-\fp{\Stot}'[\dfield, \qsource]}
	=
	\int
	\measure{\dfield} 
	\biggl(
		1 + \epsilon \fder{}{\dfield} \cdot \Psi_{\mathrm{ERG}}
		+ \epsilon \dil{\D-\Delta} \qsource \cdot \fder{}{\qsource}
		+ \epsilon V
	\biggr)
	e^{-\fp{\Stot}[\dfield, \qsource]}
	+
	\order{\epsilon^2}
,
\ee
where $ \Psi_{\mathrm{ERG}}$ can be readily deduced from~\eq{Dil-ERG}:
\be
	\Psi_{\mathrm{ERG}}
	=
	\dil{\delta} \dfield 
	+
	\dfield \cdot \ep^{-1} \cdot G
	+
	\hf \fder{}{\dfield} \cdot G
.
\ee
 That the ERG follows from a quasi-local field redefinition is repeatedly emphasised in the literature, nowhere more strongly than in~\cite{TRM+JL}. 

Of the $\dsorder{\epsilon}$ terms, the first vanishes on account of corresponding to a field redefinition and the second on account of dilatation invariance of the correlation functions. Therefore,
\be
	\ln \pf'[\qsource] = \ln \pf[\qsource] + \ln ( 1 + \epsilon V) + \order{\epsilon^2}
,
\ee
showing that the correlation functions are invariant under this procedure. Thus, for all intents and purposes, the change of coordinates~\eq{coordinate_shift} amounts to a field redefinition.

Next, consider iterating this procedure. Taking $n$ to be very large, let us write $\epsilon = a/n$. Aggregating the effects of $n$ iterations and letting $n \rightarrow \infty$ we have:
\be
	x'_\mu = \lim_{n \rightarrow \infty} (1+ a/n)^{n} x_\mu
	=
	e^{a} x_\mu = \Lambda \tilde{x}_\mu
,
\ee
where we can now interpret what we have done so far as equivalent to transferring to dimensionful variables. Continuing in this direction suppose that, in addition to shifting $x_\mu$, we also introduce a new field $\phi(x') = e^{a\fp{\delta}} \dfield(e^{-a} x')$. For the transformed action, $\Stot'[\field, \qsource]$, terms which are classically dilatation invariant have no dependence on $\Lambda$ (equivalently, $a$). The residual dependence on $\Lambda$ is captured by the dimensionful version of the ERG equation, which is essentially Polchinski's equation~\cite{Pol}. In the $\Lambda \rightarrow \infty$ limit, just the classical contributions survive. Taken as a whole, the analysis of this section justifies our earlier assertion that the boundary condition in dimension\emph{less} variables is that the action is classically dilatation invariant, up to terms which can be removed by a quasi-local field redefinition.

\providecommand{\href}[2]{#2}\begingroup\raggedright\endgroup

\end{document}